\documentclass[twocolumn,showpacs,preprintnumbers,amsmath,amssymb]{revtex4}

\usepackage{graphicx}
\usepackage{dcolumn}
\usepackage{bm}
\usepackage[english]{babel}
\usepackage{amsmath}
\usepackage{float}
\begin{document}

\title{Affine and Non-Affine Motions in Sheared Polydisperse Emulsions}

\author{J. Clara-Rahola$^1$}
\altaffiliation[Current address:  ]{Eikhonal Institute, 08750 Barcelona, Spain.}
\author{T. A. Brzinski$^1$}
\author{D. Semwogerere$^1$}
\author{K. Feitosa$^2$}
\altaffiliation[Current address:  ]{Dept. of Physics and
Astronomy, James Madison University, Harrisonburg, VA 22807, USA.}
\author{J. C. Crocker$^2$}
\author{J. Sato$^3$}
\author{V. Breedveld$^3$}
\author{Eric R. Weeks$^1$}
\email{erweeks@emory.edu}
\affiliation{$^1$Department of Physics, Emory University, 
Atlanta, GA 30322, USA}
\affiliation{$^2$Department of Chemical and Biomolecular Engineering,
University of Pennsylvania, Philadelphia, PA 19104, USA}
\affiliation{$^3$School of Chemical and Biomolecular Engineering, 
Georgia Institute of Technology,
Atlanta, GA 30332, USA}

\date{\today}

\begin{abstract}
We study dense and highly polydisperse emulsions at droplet volume
fractions $\phi \geq 0.65$.  We apply oscillatory shear and observe
droplet motion using confocal microscopy.  The presence of droplets
with sizes several times the mean size dramatically changes the
motion of smaller droplets.  Both affine and nonaffine droplet
motions are observed, with the more nonaffine motion exhibited
by the smaller droplets which are pushed around by the larger
droplets.  Droplet motions are correlated over length scales from
one to four times the mean droplet diameter, with larger length
scales corresponding to higher strain amplitudes (up to strains
of about 6\%).
\end{abstract}


\maketitle

Amorphous solids are intriguing in that they have a liquid-like
structure yet do not flow like liquids.  Window glass is
the most common example, and we have some understanding
of plastic flow of glass \cite{falk98,schuh07,bocquet09}.
Glass is not the only amorphous solid; other examples include
piles of sand, dense colloidal pastes, and shaving cream foams,
which are disordered on the scale of microns or millimeters.
Categorizing these as solid-like is reasonable as these
materials deform elastically (below a yield stress), rather
than flowing.  If a stress is applied above the yield stress,
then molecules in a glass or particles in a sand pile can
rearrange.  To make progress, most prior studies used
samples comprised of particles of one size or two similar sizes
\cite{liu96,mason97emulsions,hebraud97,falk98,yamamoto98,losert00,
petekidis02,schall07,utter08,chen10,vanhecke10,seth11,sexton11,chikkadi12}.
The picture that has developed is that the sample flows by
having small local groups of particles rearrange.
However, many natural materials of
interest are highly polydisperse, with particle sizes varying by
factors of ten or more.  The flow of such materials has been 
less widely studied \cite{princen86,saintjalmes99,jop12}.  
Differences noted from the monodisperse case include a lower
strain amplitude required for viscous flow \cite{derkach09} and
diminished sample viscosity \cite{pal96,pal10}.  The causes of
these differences are not well understood.

\begin{figure}[hbt]
	\begin{center}
		\includegraphics[width=7cm]{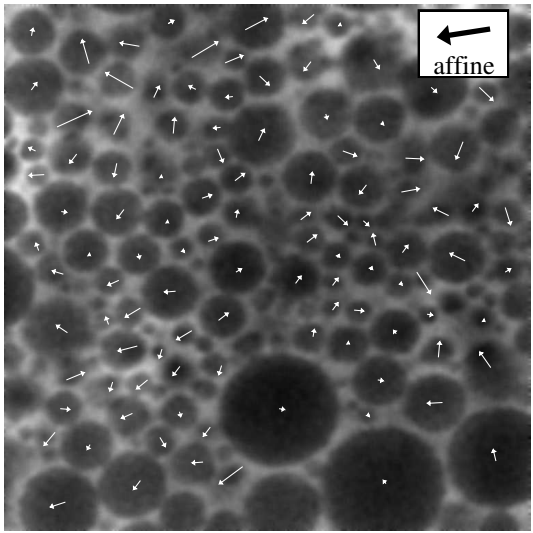}
	\end{center}
\caption{Confocal microscopy image of a polydisperse
emulsion.  The mean droplet displacement is
indicated by the large black arrow at the upper right (2.8~$\mu$m
to the left during the half-cycle); this is the affine component of
the motion.  The white arrows indicate the total displacement of
each droplet from the peak-to-peak of one half oscillation cycle,
with the mean displacement subtracted off; this is the nonaffine
component of the motion.  For easier visualization,
all arrows are twice their actual length,
including the affine displacement arrow.  The image corresponds in time to the
mid-point of this half-cycle.  The width of the image is 56~$\mu$m.
The depth is $z=24$~$\mu$m, the strain amplitude is $\gamma_{\rm
loc}=0.067$,
and the volume fraction is $\phi=0.65$.
}
\label{droplets}
\end{figure}

In this Letter, we study the shear of highly polydisperse
emulsions and show that the microscopic picture of these samples
is quite different from cases where the droplets are all
similar-sized.  Our emulsions are composed of oil droplets in water,
stabilized by a surfactant, and are at sufficiently high volume
fractions ($\phi \geq 0.65$) that the samples act as amorphous
solids \cite{saintjalmes99,derkach09,pal10}.  We subject the
samples to low amplitude oscillatory shear and follow the
droplet motion in the interior of the sample via confocal
microscopy \cite{schall07,chen10,besseling09}.  Most droplets
rearrange elastically \cite{hebraud97} and move sinusoidally.
However, these motions are not necessarily affine, as shown
in Fig.~\ref{droplets}, where the affine motion has been
subtracted off (a uniform displacement to the right for all 
droplets, indicated by the large arrow).  In particular,
our main finding is that in a highly polydisperse emulsion, the
smaller droplets frequently undergo reversible but highly nonaffine
droplet motion.  Unlike the shear of monodisperse samples
\cite{falk98,schall07,utter08,chen10,sexton11}, large
droplets allow for ``cross-talk'' between layers at different
heights which have different mean velocities.  The motions of
droplets are correlated over large length scales, up to four times
the mean droplet radius, with longer range correlations found for
higher applied strain amplitudes.  Our observations form a sharp
contrast to the localized irreversible rearrangements seen in less
polydisperse amorphous samples \cite{falk98,schall07,utter08}.

We use the shear-rupture method of Ref.~\cite{bibette96} to
create decane-in-water emulsion droplets stabilized with SDS,
skipping the fractionation step.  The continuous phase is 
a 65:35 volume ratio of water and glycerol to index match the
decane droplets.  Volume fractions are tuned to the range $0.65
\leq \phi \leq 0.85$ by centrifugation and dilution.  
Macroscopically, our samples do not flow on their own,
indicating they possess a yield stress at these volume fractions
\cite{derkach09,mason97emulsions}.

We place the samples in a parallel-plate shear cell \cite{chen10}.
The gap of the cell is fixed at $H = 200$~$\mu$m.  The lower glass
plate is fixed, and the top plate is driven sinusoidally at a
frequency $f=1$~Hz.  $f$ is chosen to be in the low-frequency limit
for this sample, where sample behavior is dominated by elastic
properties \cite{mason97emulsions}.  The amplitude is typically
$A=40$~$\mu$m, leading to a macroscopic strain amplitude of $\gamma=
A/H=0.2$.  Our experiments are conducted at volume fractions and
strain amplitudes over which the droplets remain fairly spherical.

The sample is imaged from below through the stationary plate using
a confocal microscope.  Fluorescein dye is added to the continuous
phase to visualize the droplets as shown in Fig.~\ref{droplets}.
To quantify the size distribution of our system we acquire
three-dimensional (3D) image stacks from a static sample of size
$56\times 59 \times 80$~$\mu$m$^3$.  To observe the dynamics
when sheared, we take data as rapidly as possible using
only two-dimensional (2D) images.  For the 2D experiments,
images of size $56\times 59$~$\mu$m$^{2}$ are acquired at a rate
of 90 images per second for 33~s.

\begin{figure}
	\begin{center}
		\includegraphics[width=7cm]{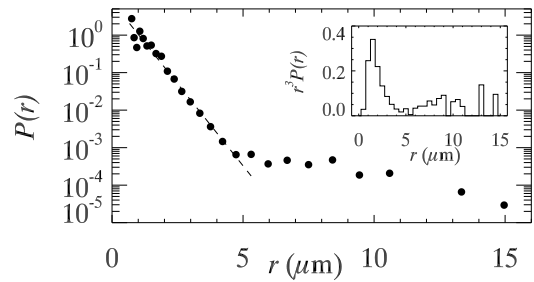}
	\end{center}
\caption{
Probability distribution of droplet sizes in a
$\phi=0.80$ sample.  Droplets with radius $r < 5$~$\mu$m are fit
to an exponential (dashed line) with decay length 0.5~$\mu$m.
The data for $r>11$~$\mu$m correspond to only three
observed droplets, thus the true shape of the distribution is
ill-defined for larger droplets.  The inset shows
the same data, with the probability weighted 
by $r^3$ and plotted on a linear scale.
}
\label{histo}
\end{figure}

Using the 3D data from static samples, we determine the 
droplet radii using custom software implementing the method of
Ref.~\cite{penfold06}.  The size distribution obtained from this
method is shown in Fig.~\ref{histo}.  The mean droplet radius is
1.2~$\mu$m, the standard deviation is 0.6~$\mu$m, and the Sauter
mean radius $r_{32} = \langle r^3 \rangle / \langle r^2 \rangle$
is 2.3~$\mu$m.  
While large droplets with $r > 5$~$\mu$m are uncommon, they account
for a nontrivial portion of the volume, as can be seen by the
volume-weighted probability distribution shown in the inset to
Fig.~\ref{histo}.

For 2D data analysis, we use a slightly different analysis
technique.  We identify droplets using the 2D-Hough transform
\cite{ioannou99}, which lets us identify the droplets' radii
and positions in each image.  From that data, we then use
conventional techniques to track their corresponding trajectories
\cite{crocker96}.  Both standard tracking and the iterative
image tracking technique described in Ref.~\cite{besseling09}
are used to reconstruct each trajectory.  Note that for each
droplet, because our observation is only in 2D, we do not know
the true radius $r_{\rm 3D}$.  However, we observe that droplets
are not distinguishable in 2D slices when they are viewed more
than $\Delta z \approx 0.6 r_{\rm 3D}$ away from their center.
This is due to the tilt of the droplet interface:  the droplet
radius changes significantly within the optical section of the 2D
confocal image ($\approx 0.6$~$\mu$m), so the edge of the droplet
is blurred in these cases and cannot be clearly determined with our
image analysis, and these droplets are not tracked.  Accordingly,
for each droplet we track, the droplet radius $r$ we apparently
observe is in the range $0.8 r_{\rm 3D} < r \leq r_{\rm 3D}$.

When an oscillatory strain is applied to the emulsion, the majority
of droplets rearrange reversibly and periodically at the driving
frequency.  The droplet-averaged displacement field is $X(z,t)
= \gamma z \sin(2 \pi f t)$ (with no motion on average in $y$
and $z$).  It is this average motion we term the ``affine motion''
in the sense that the position predicted by $X(z,t)$ is an affine
transformation of the original positions.  (This differs from
some prior work where affine motion was determined locally in
space and time \cite{utter08,falk98}.)  In Fig.~\ref{droplets}
the droplet-averaged displacement during the time interval
pictured is indicated by the large black displacement arrow.
This average motion has been subtracted off from all of the
droplets, and the remainder (the non-affine component) is indicated
by the white arrows.  The largest droplets ($r \gtrsim 8$~$\mu$m)
move sinusoidally, following $X(t,z)$, reflected by their short
displacement vectors in Fig.~\ref{droplets}.  In contrast, smaller
droplets move in a variety of directions.

This variety of displacements for the smaller droplets is 
due to the largest droplets.  At the equator of a large droplet,
it moves with the expected motion for that height $z$, that
is, $X(z,t)$.  
The droplets deform little and thus move as fairly rigid
spheres, and thus the top of a large droplet moves with an
amplitude $\gamma r$ too small relative to the expected velocity
at height $z+r$.  Likewise, at height $z-r$ the large droplet
moves faster than the mean velocity for that height.
Two large droplets that are nearby but with centers at
different $z$ do not have the same velocities, and as
they move back and forth sinusoidally, they push and pull on the
smaller droplets between them.  These smaller droplets thus have
apparently random motions.  In practice, the larger droplets are
rarer and so less likely to influence each other.  Moreover,
they move based on the average influence of the smaller droplets
surrounding them, and so their motion tends to follow the average
motion $X(z,t)$.  The contrast in motion between large and small
droplets in a highly polydisperse sample 
differs qualitatively from cases where large-scale
flows cause non-affine motion, and which typically require large
amplitude strain \cite{yamamoto98,utter08,sexton11,jop12}.

Over the 33~s movies, approximately 8\% of the droplets make
an irreversible rearrangement at some point.  This only occurs
with smaller droplets, $r < \sim 5$~$\mu$m.  Before and after
the irreversible rearrangement, the droplets move periodically.
The rarity of plastic rearrangements in our data is similar to
a prior study of a more monodisperse emulsion \cite{hebraud97}.
Our typical forcing amplitude for the shear cell ($A=40$~$\mu$m)
was chosen to limit the amount of plastic rearrangements.

At this point we only study droplets whose trajectories are
reversible (elastic) and thus periodic.  We respectively denote
as $x(t)$ and $y(t)$ the components of a trajectory parallel and
perpendicular to the shear axis at time $t$.  A least squares fit
is applied to each component with functional forms:
\begin{align}
\label{sines}
	x(t)=a_x \sin(2 \pi f t + \theta _{x})\nonumber\\
	y(t)=a_y \sin(2 \pi f t + \theta _{y})
\end{align}
using the known driving frequency $f$.  
These functional forms
provide a good fit to the particle trajectories.

We take data at several depths $z$,
relative to $z=0$ defined at the stationary bottom plate.  At each
height we compute the mean amplitude $\langle a_{x} \rangle$.
For all data, we find a linear relationship between $\langle a_x
\rangle$ and $z$, and from this linear relationship we define the
local strain $\gamma_{\rm loc}$ by $\langle a_x \rangle \sim
\gamma_{\rm loc} z$; see Supplemental Material for further
details about determining local strain.

\begin{figure}
	\begin{center}
		\includegraphics[width=8cm]{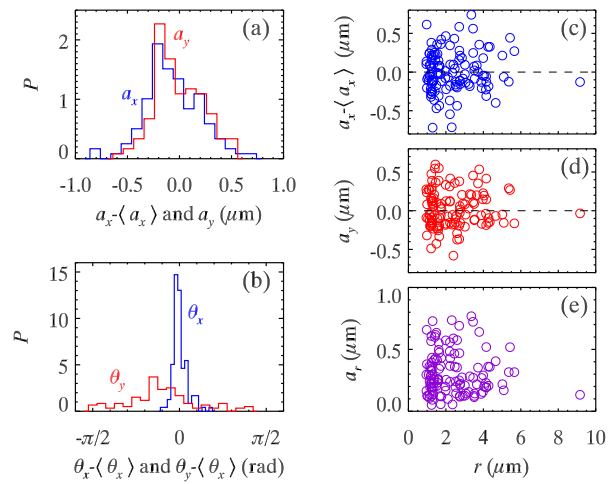}
	\end{center}
	\caption{(Color online)
(a) Probability distributions of the parallel
($a_x$) and perpendicular ($a_y$) amplitude components. (b)
Probability distributions of the phase angles.
(c-e) Scatter plots of the droplet amplitudes as a function
of their radii $r$.  (c) shows $a_x - \langle a_x \rangle$,
(d) shows $a_y$, and (e) shows $a_r \equiv \sqrt{(a_x - \langle
a_x \rangle)^2 + a_y^2}$.  The horizontal dashed lines in (c,d)
are at zero, indicating the expected value for purely affine motion.
Data are from a $\phi=0.65$ emulsion with 
$\gamma_{\rm loc}=0.070$, $z=24$~$\mu$m.	The data for
$P(a_x)$ are centered around $\langle a_x \rangle = 2.4$~$\mu$m.
$\langle \theta_x \rangle$ is arbitrary as it depends on when we
set $t=0$, although note that the distribution for $\theta_y$ is
centered around the mean value of $\theta_x$.  The lack of symmetry
in these distributions is due to the finite number of droplets.
}
\label{pdfs}
\end{figure}

The distributions of the fitting parameters $a_{x}$, $a_{y}$,
$\theta_{x}$ and $\theta_{y}$ are quite broad, as shown in
Fig.~\ref{pdfs}.  While many droplets move with the mean amplitude
$\langle a_x \rangle$ as appropriate for that height, several
have amplitudes that differ by 0.5~$\mu$m or more from the mean.
Negative values of $a_x - \langle a_x \rangle$ indicate droplets
moving with smaller amplitudes than might be expected, and likewise
positive values indicate droplets moving with larger amplitudes.
These results are equally true in the direction of the applied
strain [Fig.~\ref{pdfs}(a)] and perpendicular to this direction
[Fig.~\ref{pdfs}(b)], showing many droplets have significant
nonaffine motion.  Note that $a_y$ from our fits (Eqns.~\ref{sines})
is positive: to get values $a_y<0$, we assume that all droplets
move in phase, and so droplets with phase angles $\theta_y$
that appear $\pi$ out of phase with the dominant motion are
adjusted, $a_y \rightarrow -a_y, \theta_y \rightarrow (\theta_y
- \pi)$.  In general we find $\langle a_y \rangle \approx 0$,
as expected by symmetry.  The broad amplitude distributions we
see in Fig.~\ref{pdfs}(a,b) are qualitative similar to those seen
in Utter and Behringer's study of sheared 2D bidisperse materials
\cite{utter08}.  The widths of the amplitude distributions are in
agreement with the argument given above:  if a large droplet with
$r = 10$~$\mu$m pushes on a smaller droplet located at a height
$r$ away from the center of the large droplet, then the anomalous
motion should be $\approx \gamma_{\rm loc} r \approx 0.7$~$\mu$m
for these data.

Figure \ref{pdfs}(c-e) shows a scatter-plot of the data of
Fig.~\ref{pdfs}(a), as a function of the droplet radii $r$.  The
amplitudes associated with bigger droplets are found at the central
peaks of $P(a_x)$ and $P(a_y)$.  The outliers are more likely to
be associated with smaller sized droplets.  Figure \ref{pdfs}(e)
in particular shows the total nonaffine amplitude for each droplet,
with the larger values of this amplitude generally being seen
for smaller droplets -- although also some small droplets move
nearly affinely.

\begin{figure}
	\begin{center}
		\includegraphics[width=8cm]{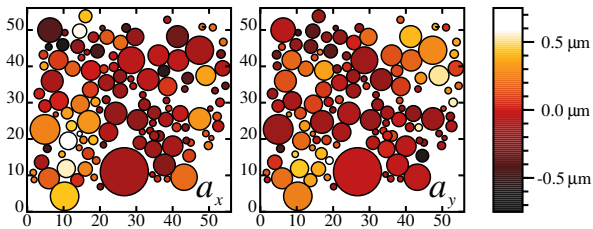}
	\end{center}
	\caption{
    \label{dropletplots}
	(Color online) Droplets which move elastically, drawn at
	their mean position.  The color code indicates the parallel
	(left picture) and perpendicular (right picture) amplitude
	of each droplet. The color bar denotes the amplitude
	scale in $\mu$m, where 0 denotes the mean amplitude for
	these data ($\langle a_x \rangle = 2.4$~$\mu$m,
    and using 0~$\mu$m for the $a_y$ data).
	The images correspond to the data shown in Fig.~\ref{pdfs}.
}
\end{figure}

To understand the spatial character of the particle behavior,
Fig.~\ref{dropletplots} shows images colored based on the values
of $a_x$ (left) and $a_y$ (right).  Droplets with similar $a_{x}$
or $a_{y}$ tend to be close together.  This is also apparent
in Fig.~\ref{droplets}, where nearby droplets have nonaffine
displacement vectors in similar directions.

\begin{figure}[tb]
	\begin{center}
		\includegraphics[width=8cm]{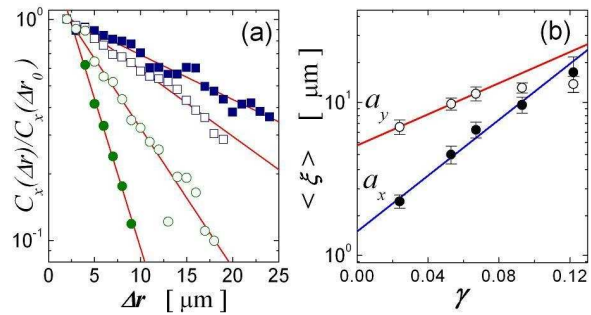}
	\end{center}
	\caption{ 
(Color online) (a) Spatial correlation functions of the
droplet amplitudes $a_x$ (solid symbols) and $a_y$ (open
symbols).  The strain amplitudes are
$\gamma_{\rm loc}=0.017$ (circles) and $\gamma_{\rm
loc}=0.085$ (squares).
The correlation functions are normalized
by their value at $\Delta r_0 = 2$~$\mu$m.
The data are from the same sample
as Figs.~\ref{pdfs} and \ref{dropletplots}.
(b) Values of the decay lengths
for this sample as a function of strain amplitude.
The uncertainty of $\gamma_{\rm loc}$ is 6\%.  The error bars are the
standard deviations found from multiple experiments.
The data shown
are taken at depth $z=24$~$\mu$m.
Lines are guides to the eye.
}
\label{corfig}
\end{figure}

To gain further insight into the spatial clustering of droplets
with similar characteristics, we identify droplet
pairs which are separated by a surface-to-surface distance
$\Delta r$, that is, droplets separated by a center-to-center
distance $r_{12}=r_{1}+r_{2}+\Delta r$, where $r_{i}$ is the
radius of droplet $i$ in the 2D image. Then, we compute a spatial
correlation function for the amplitude $a_x$ (and similar for $a_y$):
\begin{equation}
C(\Delta r) = \frac{1}{N(\Delta r)}
\sum_{\langle i, j\rangle}\frac{(a_{x_{i}}-\langle
a_x \rangle)(a_{x_{j}}-\langle
a_x\rangle)}{\sigma_{a_x}^2}
\end{equation}
where $N(\Delta r)$ is the number of neighboring droplets, and
$a_{x_i}$ and $a_{x_j}$ are the $x$-amplitudes of droplets $i$
and $j$. The average $\langle a_x \rangle$ is for all droplets
comprised by $N(\Delta r)$ and $\sigma_{a_x}$ corresponds to
the variance of the  distribution of $a_{x}$.  The choice of using
the surface-to-surface distances rather than center-to-center is
perhaps not obvious.  However, each individual droplet moves as a
solid object, thus completely correlated with itself (distances
$r_{12} < r_1$).  Examining the surface-to-surface motion lets us
avoid considering this artificially correlated solid-body motion,
and instead probe the properties of the effective medium between
droplets.  If we consider center-to-center separations, the results
are noisy and do not have a simple dependence on the distance.

Figure \ref{corfig}(a) shows these correlation functions for $a_x$
(solid symbols) and $a_y$ (open symbols), for one sample at two
different strain amplitudes.  The correlation functions exhibit
exponential decay with decay lengths in the range of 8-15~$\mu$m;
the decay lengths for different samples are provided in the
Supplemental Material and do not vary systematically with
volume fraction.
These lengths are comparable to the sizes of the larger droplets
in the sample.  Figure \ref{corfig}(a) shows that correlations
in the larger strain case (squares) decay slower than than in
the small strain case (circles).  We find $\xi_x$ and $\xi_y$
depend on $\gamma_{\rm loc}$ as shown in Fig.~\ref{corfig}(b)
for this $\phi=0.65$ sample.  

We have studied dense polydisperse emulsions, and observed highly
complex droplet motion when our samples are sinusoidally sheared.
Most droplets move periodically, but different droplets have
different amplitudes and phases.  Large
droplets push small droplets out of their way, although nearly
all of this motion is reversible.  In fact, a key point is that
the complex droplet motions occur at low strain amplitudes where
the behavior is elastic, rather than requiring large amplitude
plastic flow.
We find length
scales over which droplet motions are correlated.  These length
scales range from 1 to 4 times the mean droplet diameter, with
the largest values found for the highest strains ($\gamma_{\rm
loc} \sim 0.08$).
Overall, our results suggest that the flow 
of highly polydisperse systems is richer than that of
monodisperse samples.
Theoretical descriptions derived for less polydisperse systems
will likely not apply or need to be modified \cite{liu96,falk98}.
Preliminary observations of steadily sheared polydisperse emulsions
suggest that these results carry over in a qualitative respect,
with the largest droplets moving in straightforward fashion under
steady shear, and the smallest droplets moving in highly variable
trajectories.

We thank G.~W.~Baxter, R.~Besseling, C.~Crane, R.~Gonzalez,
C.~Hollinger, and W.~C.~K.~Poon for helpful discussions.
Funding from the National Science Foundation (grants DMR-0603055
and DMR-1336401), the Petroleum Research Fund (administered by
the American Chemical Society, grant 47970-AC9), and the Swiss
National Foundation (grant PBFR2-116930) is gratefully acknowledged.

\clearpage

\section*{Affine and Non-Affine Motions in Sheared Polydisperse
Emulsions:  Supplemental Material}

Prior to our experiments, the samples are gently stirred to
prevent any size segregation due to sedimentation, although for
our high volume fractions sedimentation and size segregation are
exceedingly slow.  To try to ensure that the sample does not slip,
where the sample comes into contact with the glass plates, we add
a coating of ScotchGard (3M).  The sample wets the ScotchGard,
pinning the sample to the coated region and ensuring a no-slip
boundary condition at each plate.  We do not observe the behavior of
purely slipping emulsions in any of our experiments [S1].

We study the sample at depths $z=24, 36, 48,$ and 60~$\mu$m,
relative to $z=0$ defined at the stationary bottom plate.  At each
height we compute the mean amplitude $\langle a_{x} \rangle$.
We find $\langle a_{x} \rangle\sim \gamma_{\rm loc} \cdot(z - z_0)$
where $\gamma_{\rm loc}$ is the local strain amplitude and $z_0$
is a slip length.  $\langle a_x \rangle$ does not extrapolate to 0
at $z=0$, but rather at negative values ranging from $z_0 = -8$ to
$-14$~$\mu$m.  The local strain is always smaller than the applied
strain.  This suggests that the emulsion partially slips at the
top plate, or possibly has a shear band somewhere where the local
strain is significantly higher.  Unfortunately, our confocal could
not image deeply enough to observe the behavior at the top plate.
We emphasize that the mean strain is uniform throughout within the
the volume we image.  See the Table for further information about
each particular experiment.

Given the dependence of the decay length $\xi$ on $\gamma_{\rm
loc}$ (as discussed in the main text), it suggests that perhaps
the variation we see for samples with different $\phi$ in the
Table is perhaps more due to the variability in $\gamma_{\rm loc}$
for those data.

\begin{table}[bh]
\begin{tabular}{ccccc}
$\phi$ & $\gamma_{\rm loc}$ & $z_0 [\mu m]$
& $\xi_{x} [\mu m]$ & $\xi_{y} [\mu m]$ \\
\hline
\hline
%phi   gamma    z0        xi_x   xi_y 
0.65 & 0.070 &  -9.0    &  9.4 & 10.3  \\
0.70 & 0.102 & -13.8    & 14.7 &  9.1  \\
0.75 & 0.054 &  -9.2    &  8.0 &  9.6  \\
0.85 & 0.070 &  -8.8    &  8.3 & 11.7  \\
\hline
\end{tabular}
\caption{Volume fraction $\phi$, observed local strain amplitudes 
$\gamma_{\rm loc}$ for different
volume fractions $\phi$, observed slip length $z_0$,
and the characteristic length scales $\xi_x$ and $\xi_y$.
The macroscopic applied strain amplitude is $0.20$ for all
experiments.  The uncertainties for all listed lengths are $\pm
0.3$~$\mu$m.
}
\end{table}

\section*{Supplemental Reference}

[S1] S.~P.~Meeker, R.~T.~Bonnecaze, and M.~Cloitre,
Phys. Rev.~Lett.~{\bf 92}, 198302 (2004).


\begin{thebibliography}{27}
\expandafter\ifx\csname natexlab\endcsname\relax\def\natexlab#1{#1}\fi
\expandafter\ifx\csname bibnamefont\endcsname\relax
  \def\bibnamefont#1{#1}\fi
\expandafter\ifx\csname bibfnamefont\endcsname\relax
  \def\bibfnamefont#1{#1}\fi
\expandafter\ifx\csname citenamefont\endcsname\relax
  \def\citenamefont#1{#1}\fi
\expandafter\ifx\csname url\endcsname\relax
  \def\url#1{\texttt{#1}}\fi
\expandafter\ifx\csname urlprefix\endcsname\relax\def\urlprefix{URL }\fi
\providecommand{\bibinfo}[2]{#2}
\providecommand{\eprint}[2][]{\url{#2}}

\bibitem[{\citenamefont{Falk and Langer}(1998)}]{falk98}
\bibinfo{author}{\bibfnamefont{M.~L.} \bibnamefont{Falk}} \bibnamefont{and}
  \bibinfo{author}{\bibfnamefont{J.~S.} \bibnamefont{Langer}},
  \bibinfo{journal}{Phys. Rev. E} \textbf{\bibinfo{volume}{57}},
  \bibinfo{pages}{7192} (\bibinfo{year}{1998}).

\bibitem[{\citenamefont{Schuh et~al.}(2007)\citenamefont{Schuh, Hufnagel, and
  Ramamurty}}]{schuh07}
\bibinfo{author}{\bibfnamefont{C.~A.} \bibnamefont{Schuh}},
  \bibinfo{author}{\bibfnamefont{T.~C.} \bibnamefont{Hufnagel}},
  \bibnamefont{and}
  \bibinfo{author}{\bibfnamefont{U.}~\bibnamefont{Ramamurty}},
  \bibinfo{journal}{Acta Materialia} \textbf{\bibinfo{volume}{55}},
  \bibinfo{pages}{4067} (\bibinfo{year}{2007}).

\bibitem[{\citenamefont{Bocquet et~al.}(2009)\citenamefont{Bocquet, Colin, and
  Ajdari}}]{bocquet09}
\bibinfo{author}{\bibfnamefont{L.}~\bibnamefont{Bocquet}},
  \bibinfo{author}{\bibfnamefont{A.}~\bibnamefont{Colin}}, \bibnamefont{and}
  \bibinfo{author}{\bibfnamefont{A.}~\bibnamefont{Ajdari}},
  \bibinfo{journal}{Phys. Rev. Lett.} \textbf{\bibinfo{volume}{103}},
  \bibinfo{pages}{036001} (\bibinfo{year}{2009}).

\bibitem[{\citenamefont{Liu et~al.}(1996)\citenamefont{Liu, Ramaswamy, Mason,
  Gang, and Weitz}}]{liu96}
\bibinfo{author}{\bibfnamefont{A.~J.} \bibnamefont{Liu}},
  \bibinfo{author}{\bibfnamefont{S.}~\bibnamefont{Ramaswamy}},
  \bibinfo{author}{\bibfnamefont{T.~G.} \bibnamefont{Mason}},
  \bibinfo{author}{\bibfnamefont{H.}~\bibnamefont{Gang}}, \bibnamefont{and}
  \bibinfo{author}{\bibfnamefont{D.~A.} \bibnamefont{Weitz}},
  \bibinfo{journal}{Phys. Rev. Lett.} \textbf{\bibinfo{volume}{76}},
  \bibinfo{pages}{3017} (\bibinfo{year}{1996}).

\bibitem[{\citenamefont{Mason et~al.}(1997)\citenamefont{Mason, Lacasse, Grest,
  Levine, Bibette, and Weitz}}]{mason97emulsions}
\bibinfo{author}{\bibfnamefont{T.~G.} \bibnamefont{Mason}},
  \bibinfo{author}{\bibfnamefont{M.-D.} \bibnamefont{Lacasse}},
  \bibinfo{author}{\bibfnamefont{G.~S.} \bibnamefont{Grest}},
  \bibinfo{author}{\bibfnamefont{D.}~\bibnamefont{Levine}},
  \bibinfo{author}{\bibfnamefont{J.}~\bibnamefont{Bibette}}, \bibnamefont{and}
  \bibinfo{author}{\bibfnamefont{D.~A.} \bibnamefont{Weitz}},
  \bibinfo{journal}{Phys. Rev. E} \textbf{\bibinfo{volume}{56}},
  \bibinfo{pages}{3150} (\bibinfo{year}{1997}).

\bibitem[{\citenamefont{H\'{e}braud et~al.}(1997)\citenamefont{H\'{e}braud,
  Lequeux, Munch, and Pine}}]{hebraud97}
\bibinfo{author}{\bibfnamefont{P.}~\bibnamefont{H\'{e}braud}},
  \bibinfo{author}{\bibfnamefont{F.}~\bibnamefont{Lequeux}},
  \bibinfo{author}{\bibfnamefont{J.}~\bibnamefont{Munch}}, \bibnamefont{and}
  \bibinfo{author}{\bibfnamefont{D.}~\bibnamefont{Pine}},
  \bibinfo{journal}{Phys. Rev. Lett.} \textbf{\bibinfo{volume}{78}},
  \bibinfo{pages}{4657} (\bibinfo{year}{1997}).

\bibitem[{\citenamefont{Yamamoto and Onuki}(1998)}]{yamamoto98}
\bibinfo{author}{\bibfnamefont{R.}~\bibnamefont{Yamamoto}} \bibnamefont{and}
  \bibinfo{author}{\bibfnamefont{A.}~\bibnamefont{Onuki}},
  \bibinfo{journal}{Phys. Rev. E} \textbf{\bibinfo{volume}{58}},
  \bibinfo{pages}{3515} (\bibinfo{year}{1998}).

\bibitem[{\citenamefont{Losert et~al.}(2000)\citenamefont{Losert, Bocquet,
  Lubensky, and Gollub}}]{losert00}
\bibinfo{author}{\bibfnamefont{W.}~\bibnamefont{Losert}},
  \bibinfo{author}{\bibfnamefont{L.}~\bibnamefont{Bocquet}},
  \bibinfo{author}{\bibfnamefont{T.~C.} \bibnamefont{Lubensky}},
  \bibnamefont{and} \bibinfo{author}{\bibfnamefont{J.~P.}
  \bibnamefont{Gollub}}, \bibinfo{journal}{Phys. Rev. Lett.}
  \textbf{\bibinfo{volume}{85}}, \bibinfo{pages}{1428} (\bibinfo{year}{2000}).

\bibitem[{\citenamefont{Petekidis et~al.}(2002)\citenamefont{Petekidis,
  Moussa\"{i}d, and Pusey}}]{petekidis02}
\bibinfo{author}{\bibfnamefont{G.}~\bibnamefont{Petekidis}},
  \bibinfo{author}{\bibfnamefont{A.}~\bibnamefont{Moussa\"{i}d}},
  \bibnamefont{and} \bibinfo{author}{\bibfnamefont{P.~N.} \bibnamefont{Pusey}},
  \bibinfo{journal}{Phys. Rev. E} \textbf{\bibinfo{volume}{66}},
  \bibinfo{pages}{051402} (\bibinfo{year}{2002}).

\bibitem[{\citenamefont{Schall et~al.}(2007)\citenamefont{Schall, Weitz, and
  Spaepen}}]{schall07}
\bibinfo{author}{\bibfnamefont{P.}~\bibnamefont{Schall}},
  \bibinfo{author}{\bibfnamefont{D.~A.} \bibnamefont{Weitz}}, \bibnamefont{and}
  \bibinfo{author}{\bibfnamefont{F.}~\bibnamefont{Spaepen}},
  \bibinfo{journal}{Science} \textbf{\bibinfo{volume}{318}},
  \bibinfo{pages}{1895} (\bibinfo{year}{2007}).

\bibitem[{\citenamefont{Utter and Behringer}(2008)}]{utter08}
\bibinfo{author}{\bibfnamefont{B.}~\bibnamefont{Utter}} \bibnamefont{and}
  \bibinfo{author}{\bibfnamefont{R.~P.} \bibnamefont{Behringer}},
  \bibinfo{journal}{Phys. Rev. Lett.} \textbf{\bibinfo{volume}{100}},
  \bibinfo{pages}{208302} (\bibinfo{year}{2008}).

\bibitem[{\citenamefont{Chen et~al.}(2010)\citenamefont{Chen, Semwogerere,
  Sato, Breedveld, and Weeks}}]{chen10}
\bibinfo{author}{\bibfnamefont{D.}~\bibnamefont{Chen}},
  \bibinfo{author}{\bibfnamefont{D.}~\bibnamefont{Semwogerere}},
  \bibinfo{author}{\bibfnamefont{J.}~\bibnamefont{Sato}},
  \bibinfo{author}{\bibfnamefont{V.}~\bibnamefont{Breedveld}},
  \bibnamefont{and} \bibinfo{author}{\bibfnamefont{E.~R.} \bibnamefont{Weeks}},
  \bibinfo{journal}{Phys. Rev. E} \textbf{\bibinfo{volume}{81}},
  \bibinfo{pages}{011403} (\bibinfo{year}{2010}).

\bibitem[{\citenamefont{Hecke}(2010)}]{vanhecke10}
\bibinfo{author}{\bibfnamefont{M.~V.} \bibnamefont{Hecke}},
  \bibinfo{journal}{J. Phys.: Cond. Matt.r} \textbf{\bibinfo{volume}{22}},
  \bibinfo{pages}{033101} (\bibinfo{year}{2010}).

\bibitem[{\citenamefont{Seth et~al.}(2011)\citenamefont{Seth, Mohan,
  Locatelli-Champagne, Cloitre, and Bonnecaze}}]{seth11}
\bibinfo{author}{\bibfnamefont{J.~R.} \bibnamefont{Seth}},
  \bibinfo{author}{\bibfnamefont{L.}~\bibnamefont{Mohan}},
  \bibinfo{author}{\bibfnamefont{C.~A.} \bibnamefont{Locatelli-Champagne}},
  \bibinfo{author}{\bibfnamefont{M.}~\bibnamefont{Cloitre}}, \bibnamefont{and}
  \bibinfo{author}{\bibfnamefont{R.~T.} \bibnamefont{Bonnecaze}},
  \bibinfo{journal}{Nature Materials} \textbf{\bibinfo{volume}{10}},
  \bibinfo{pages}{838} (\bibinfo{year}{2011}).

\bibitem[{\citenamefont{Sexton et~al.}(2011)\citenamefont{Sexton, M{\"o}bius,
  and Hutzler}}]{sexton11}
\bibinfo{author}{\bibfnamefont{M.~B.} \bibnamefont{Sexton}},
  \bibinfo{author}{\bibfnamefont{M.~E.} \bibnamefont{M{\"o}bius}},
  \bibnamefont{and} \bibinfo{author}{\bibfnamefont{S.}~\bibnamefont{Hutzler}},
  \bibinfo{journal}{Soft Matter} \textbf{\bibinfo{volume}{7}},
  \bibinfo{pages}{11252} (\bibinfo{year}{2011}).

\bibitem[{\citenamefont{Chikkadi and Schall}(2012)}]{chikkadi12}
\bibinfo{author}{\bibfnamefont{V.}~\bibnamefont{Chikkadi}} \bibnamefont{and}
  \bibinfo{author}{\bibfnamefont{P.}~\bibnamefont{Schall}},
  \bibinfo{journal}{Phys. Rev. E} \textbf{\bibinfo{volume}{85}},
  \bibinfo{pages}{031402} (\bibinfo{year}{2012}).

\bibitem[{\citenamefont{Princen}(1986)}]{princen86}
\bibinfo{author}{\bibfnamefont{H.}~\bibnamefont{Princen}}, \bibinfo{journal}{J.
  Colloid Interface Sci.} \textbf{\bibinfo{volume}{112}}, \bibinfo{pages}{427}
  (\bibinfo{year}{1986}).

\bibitem[{\citenamefont{Saint-Jalmes and Durian}(1999)}]{saintjalmes99}
\bibinfo{author}{\bibfnamefont{A.}~\bibnamefont{Saint-Jalmes}}
  \bibnamefont{and} \bibinfo{author}{\bibfnamefont{D.~J.}
  \bibnamefont{Durian}}, \bibinfo{journal}{J. Rheo.}
  \textbf{\bibinfo{volume}{43}}, \bibinfo{pages}{1411} (\bibinfo{year}{1999}).

\bibitem[{\citenamefont{Jop et~al.}(2012)\citenamefont{Jop, Mansard, Chaudhuri,
  Bocquet, and Colin}}]{jop12}
\bibinfo{author}{\bibfnamefont{P.}~\bibnamefont{Jop}},
  \bibinfo{author}{\bibfnamefont{V.}~\bibnamefont{Mansard}},
  \bibinfo{author}{\bibfnamefont{P.}~\bibnamefont{Chaudhuri}},
  \bibinfo{author}{\bibfnamefont{L.}~\bibnamefont{Bocquet}}, \bibnamefont{and}
  \bibinfo{author}{\bibfnamefont{A.}~\bibnamefont{Colin}},
  \bibinfo{journal}{Phys. Rev. Lett.} \textbf{\bibinfo{volume}{108}},
  \bibinfo{pages}{148301} (\bibinfo{year}{2012}).

\bibitem[{\citenamefont{Derkach}(2009)}]{derkach09}
\bibinfo{author}{\bibfnamefont{S.~R.} \bibnamefont{Derkach}},
  \bibinfo{journal}{Adv. Coll. Int. Sci.} \textbf{\bibinfo{volume}{151}},
  \bibinfo{pages}{1} (\bibinfo{year}{2009}).

\bibitem[{\citenamefont{Pal}(1996)}]{pal96}
\bibinfo{author}{\bibfnamefont{R.}~\bibnamefont{Pal}}, \bibinfo{journal}{AIChE
  J.} \textbf{\bibinfo{volume}{42}}, \bibinfo{pages}{3181}
  (\bibinfo{year}{1996}).

\bibitem[{\citenamefont{Pal}(2011)}]{pal10}
\bibinfo{author}{\bibfnamefont{R.}~\bibnamefont{Pal}}, \bibinfo{journal}{Curr.
  Op. Coll. Int. Sci.} \textbf{\bibinfo{volume}{16}}, \bibinfo{pages}{41}
  (\bibinfo{year}{2011}).

\bibitem[{\citenamefont{Besseling et~al.}(2009)\citenamefont{Besseling, Isa,
  Weeks, and Poon}}]{besseling09}
\bibinfo{author}{\bibfnamefont{R.}~\bibnamefont{Besseling}},
  \bibinfo{author}{\bibfnamefont{L.}~\bibnamefont{Isa}},
  \bibinfo{author}{\bibfnamefont{E.~R.} \bibnamefont{Weeks}}, \bibnamefont{and}
  \bibinfo{author}{\bibfnamefont{W.~C.~K.} \bibnamefont{Poon}},
  \bibinfo{journal}{Adv. Coll. Int. Sci.} \textbf{\bibinfo{volume}{146}},
  \bibinfo{pages}{1} (\bibinfo{year}{2009}).

\bibitem[{\citenamefont{Mason and Bibette}(1996)}]{bibette96}
\bibinfo{author}{\bibfnamefont{T.~G.} \bibnamefont{Mason}} \bibnamefont{and}
  \bibinfo{author}{\bibfnamefont{J.}~\bibnamefont{Bibette}},
  \bibinfo{journal}{Phys. Rev. Lett.} \textbf{\bibinfo{volume}{77}},
  \bibinfo{pages}{3481} (\bibinfo{year}{1996}).

\bibitem[{\citenamefont{Penfold et~al.}(2006)\citenamefont{Penfold, Watson,
  Mackie, and Hibberd}}]{penfold06}
\bibinfo{author}{\bibfnamefont{R.}~\bibnamefont{Penfold}},
  \bibinfo{author}{\bibfnamefont{A.~D.} \bibnamefont{Watson}},
  \bibinfo{author}{\bibfnamefont{A.~R.} \bibnamefont{Mackie}},
  \bibnamefont{and} \bibinfo{author}{\bibfnamefont{D.~J.}
  \bibnamefont{Hibberd}}, \bibinfo{journal}{Langmuir}
  \textbf{\bibinfo{volume}{22}}, \bibinfo{pages}{2005} (\bibinfo{year}{2006}).

\bibitem[{\citenamefont{Ioannou et~al.}(1999)\citenamefont{Ioannou, Huda, and
  Laine}}]{ioannou99}
\bibinfo{author}{\bibfnamefont{D.}~\bibnamefont{Ioannou}},
  \bibinfo{author}{\bibfnamefont{W.}~\bibnamefont{Huda}}, \bibnamefont{and}
  \bibinfo{author}{\bibfnamefont{A.~F.} \bibnamefont{Laine}},
  \bibinfo{journal}{Image and Vision Computing} \textbf{\bibinfo{volume}{17}},
  \bibinfo{pages}{15} (\bibinfo{year}{1999}).

\bibitem[{\citenamefont{Crocker and Grier}(1996)}]{crocker96}
\bibinfo{author}{\bibfnamefont{J.~C.} \bibnamefont{Crocker}} \bibnamefont{and}
  \bibinfo{author}{\bibfnamefont{D.~G.} \bibnamefont{Grier}},
  \bibinfo{journal}{J. Colloid Interface Sci.} \textbf{\bibinfo{volume}{179}},
  \bibinfo{pages}{298} (\bibinfo{year}{1996}).

\end{thebibliography}
\end{document}